\documentstyle[twocolumn,aps,epsf]{revtex}

\begin{document}

\title{Long-range effects on optical absorption in quasiperiodic
lattices}

\author{Francisco Dom\'{\i}nguez-Adame${\dag}$}

\address{Departamento de F\'{\i}sica de Materiales,
Facultad de F\'{\i}sicas, Universidad Complutense,\\
E-28040 Madrid, Spain \\
\bigskip}

\author{\small\parbox{14cm}{\small
We consider exciton optical absorption in quasiperiodic lattices,
focusing our attention on the Fibonacci case as typical example. The
absorption spectrum is evaluated by solving numerically the equation of
motion of the Frenkel-exciton problem on the lattice, in which on-site
energies take on two values according to the Fibonacci sequence. We find
that the quasiperiodic order causes the occurrence of well-defined
characteristic features in the absorption spectra. We also develop an
analytical method that relates satellite lines with the Fourier pattern
of the lattice. Our predictions can be used to determine experimentally
the long-range quasiperiodic order from optical measurements.\\[3pt]
}
}\address{}\maketitle

\section{Introduction}

Linear lattices described by means of the binary Fibonacci sequence have
been regarded as one-dimensional (1D) quasicrystals and, consequently,
they have been the subject of intensive theoretical studies.  These
systems lack translational symmetry but, unlike disordered lattices,
display long-range order.  The corresponding electronic states present
rather unusual properties like fractal energy spectra and self-similar
wave functions.  From the experimental viewpoint, several electronic
properties of solids can be inferred from optical characterization
techniques like optical absorption, photoluminescence and fluorescence
after pulse excitation.  Then, a complete understanding of the interplay
between the electronic properties and the underlying quasiperiodic order
requires a detailed analysis of optical processes.  This reason has
motivated various works dealing with optical properties of Fibonacci
systems, mainly devoted to Fibonacci semiconductor superlattices
\cite{DasSarma,SSC,Tuet,Munzar}.

In previous works we have focused our attention on excitons in Fibonacci
lattices \cite{OPTFIBO,masterfib}. In particular, we have numerically
evaluated optical absorption spectra due to Frenkel excitons in
self-similar aperiodic lattices (Fibonacci and Thue-Morse)
\cite{OPTFIBO}.  We found several characteristic lines specific of each
aperiodic system that are not present in periodic or random Frenkel
lattices, thus being an adequate way for determining the particular
structural order of the system from experiments.  When the two kind of
optical centers composing the aperiodic lattice are very different, the
origin of all the absorption lines can be successfully assigned by
considering the aperiodic lattices as composed of two-center blocks, as
much as in the same spirit of the renormalization group concepts
developed in the last few years \cite{Niu,Liu}.  This approach neglects
the long-range order present in the lattice and only takes into account
the short-range interaction between nearest-neighbor optical centers.
These results lead, in a natural way, to the question whether long-range
quasiperiodic order can also be characterized by means of optical
absorption spectra.  It is clear that we must search for specific lines
due to the long-range order in the opposite limit, namely when the two
kind of optical centers are of similar nature.  We will show below that
this is indeed the case.

In this work we investigate optical absorption due to Frenkel excitons
in 1D binary systems arranged according to the Fibonacci sequence.  We
make use of a general treatment which allows us to study the dynamics of
Frenkel excitons in these lattices, solve the microscopic equations of
motion and find the optical absorption spectrum.  The main conclusions
of this paper are twofold.  First, we show that Frenkel excitons in
Fibonacci lattices with similar optical centers lead to absorption lines
specific of this kind of ordering and, consequently, not present in
periodic or disordered lattices.  Second, and most important, by means
of an analytical approach we are able to explain the origin of these
characteristics lines, which are related to the Fourier transform of the
lattice.  Thus we successfully relate an optical property (optical
spectrum) with an structural one (topology of the lattice).

\section{Physical Model and Theory}

We consider a system of $N$ optically active, two-level centers,
occupying positions ${\bf r}_n$ on a 1D regular lattice with spacing
unity.  Therefore, the effective Frenkel Hamiltonian describing this
system can be written (we use units such that $\hbar=1$)
\begin{equation}
{\cal H}= \sum_{n}\> V_{n} a_{n}^{\dag}a_{n} +
\sum_{l\neq n}\>J_{nl}a_{n}^{\dag}a_{l}.
\label{Hamiltonian}
\end{equation}
Here $a_{n}^{\dag}$ and $a_{n}$ creates and annihilates an electronic
excitation of energy $V_n$ at site $n$, respectively. $J_{nl}$ ($n\neq
l$) is the intersite interaction of dipole origin between centers $n$
and $l$, assumed to be of the form $J_{nl}=-J|{\bf r}_n-{\bf
r}_l|^{-3}$, where $J$ is the coupling between nearest-neighbor centers.
Since $J_{nl}$ is a rapidly decreasing function of the distance between
centers, we will omit interactions beyond nearest-neighbors.

In this work we will be concerned with the Frenkel exciton dynamics in
systems presenting a quasiperiodic distribution of optical centers.  The
Fibonacci lattice is the archetypal example of deterministic and
quasiperiodically ordered structure.  Any arbitrary Fibonacci system
presents two kind of building blocks.  In our case, we choose those
blocks as individual two-level centers with on-site energies $V_A$ and
$V_B$.  The Fibonacci arrangement can be generated by the substitution
rule $A\rightarrow AB$, $B\rightarrow A$. In this way, finite and
self-similar quasiperiodic lattices are obtained by $n$ successive
applications of the substitution rule.  The $n$th generation lattice
will have $N=F_n$ elements, where $F_n$ denotes a Fibonacci number.
Such numbers are generated from the recurrence law $F_n=F_{n-1} +
F_{n-2}$ starting with $F_0=F_1=1$; as $n$ increases the ratio
$F_{n-1}/F_n$ converges toward $\tau=(\sqrt{5}-1)/2=0.618\ldots$, an
irrational number which is known as the inverse golden mean.  Therefore,
lattice sites are arranged according to the Fibonacci sequence
$V_A\,V_B\,V_A\,V_A\,V_B\,V_A\,V_B\,V_A\ldots$ where the fraction of
$B$-centers is $c\sim 1-\tau$.

Having presented our model we now briefly describe the method we have
used to calculate the absorption spectra.  The line shape $I(E)$ of an
optical-absorption process in which a single exciton is created in a
lattice with $N$ sites can be obtained as follows \cite{Huber}.  Let us
consider the total dipole moment operator ${\cal D}=\sum_n\>
(a_{n}^{\dag}+a_{n})$, where the dipole moment of each center is taken
to be unity.  Here we are restricting to the case of systems whose
length is much smaller than the optical wavelength.  Denoting by $|k
\rangle$ and $E_k$ the eigenvectors and eigenvalues of the Hamiltonian
$\cal H$, respectively, the absorption line-shape is given by
\begin{equation}
I(E)={1\over N} \sum_{k}\> \left| \langle k| {\cal D} |
\mbox{vac} \rangle \right|^{2} \delta (E-E_k),
\label{primero}
\end{equation}
where $|\mbox{vac}\rangle$ denotes the exciton vacuum.  In practice one
considers a broadened $\delta$-function to take into account the
instrumental resolution function.  Therefore we will replace the
$\delta$-function by a Lorentzian function of half-width $\alpha$,
hereafter denoted by $\delta_{\alpha}$.

A reliable method for determining $I(E)$ numerically involves the
consideration of the correlation functions \cite{Huber}
\begin{equation}
G_{n}(t)=\sum_{l}\>\langle \mbox{vac}|a_{n}(t)a_{l}^{\dag}|
\mbox{vac}\rangle,
\label{G}
\end{equation}
where $a_{n}(t)= \exp (i{\cal H}t) a_{n} \exp (-i{\cal H}t)$ is the
annihilation operator in the Heisenberg representation.  The function
$G_{n}(t)$ obeys the equation of motion
\begin{equation}
i{d\over dt} G_{n}(t) = \sum_{l}\> H_{nl} G_{l}(t),
\label{motion}
\end{equation}
with the initial condition $G_{n}(0)=1$.  The diagonal elements of the
tridiagonal matrix $H_{nl}$ are $V_{n}$ whereas off-diagonal elements
are simply given by $-J$.  The microscopic equation of motion is a
discrete Schr\"odinger-like equation on a lattice and standard numerical
techniques may be applied to obtain the solution.  Once these equations
of motion are solved, the line shape is found from the following
expression
\begin{equation}
I(E)=-\,{2\over \pi N} \int_0^\infty\> dt\, e^{-\alpha t} \sin (Et)
\mbox{Im}\left( \sum_{n} G_{n} (t) \right),
\label{spectrum}
\end{equation}
where the factor $\exp(-\alpha t)$ takes into account the broadening
due to the instrumental resolution function of half-width $\alpha$.

\begin{figure}

\centerline{\epsfxsize=8.0truecm \epsffile{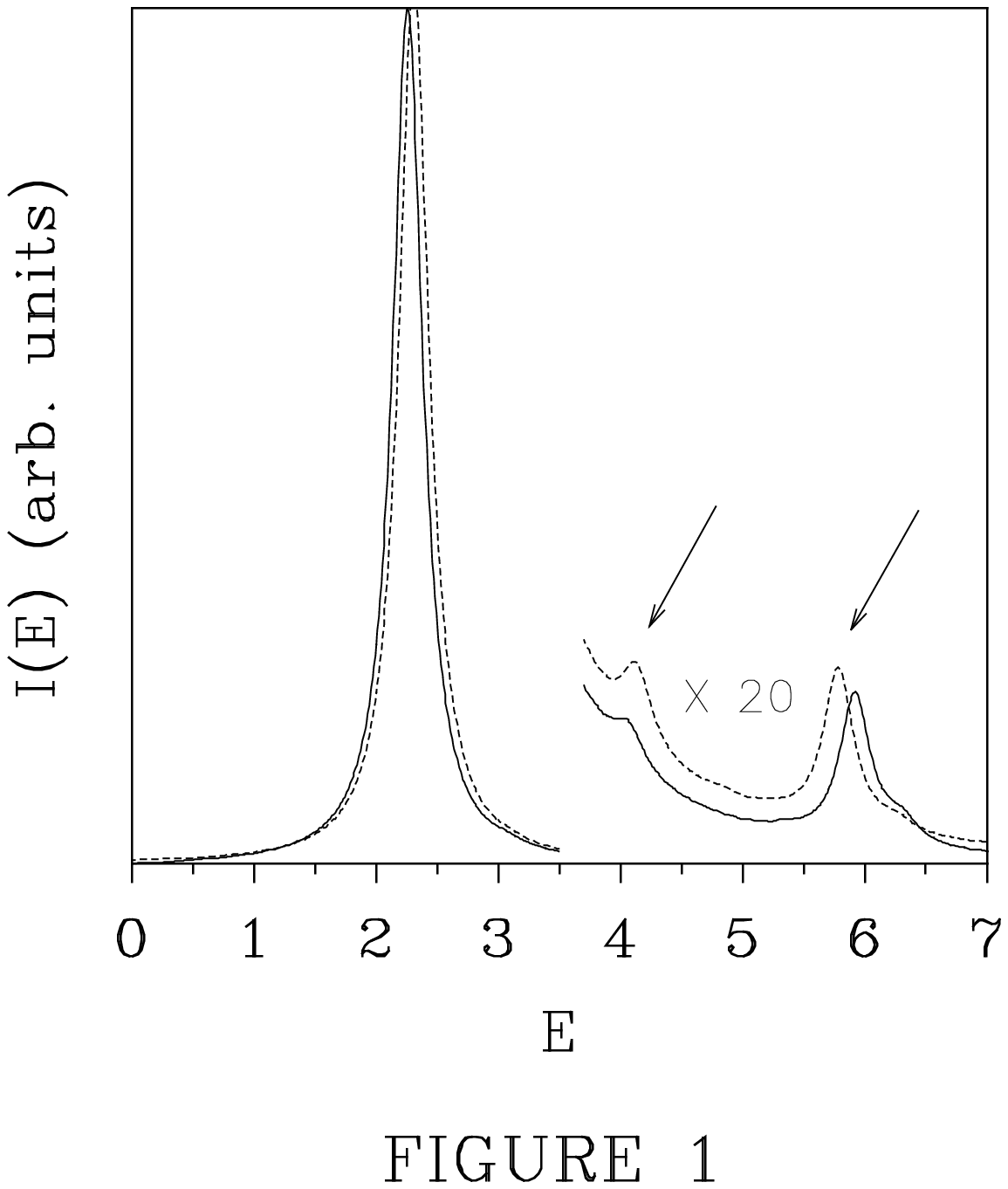}}

\caption{Absorption spectra for a Fibonacci lattice with
$N=F_{12}=233$ active centers and on-site energies $V_A=4.0$ and
$V_B=4.8$. Solid line indicates numerical result and dashed line
the analytical approach.}

\label{fig1}

\end{figure}

\section{Numerical Results}

We have solved numerically the equation of motion (\ref{motion}) using
an implicit (Crank-Nicholson) integration scheme \cite{Recipes}.  In the
remainder of the paper, energy will be measured in units of $J$ whereas
time will be expressed in units of $J^{-1}$.  The energy and time scales
can deduced from the experiment since the exciton bandwidth is $4J$.
Fibonacci lattices are generated using the inflation rules discussed
above.  In order to minimize end effects, spatial periodic boundary
conditions are introduced in all cases.  We have checked that the
position and the strength of all lines of the spectra are independent of
the system size.  Hereafter we will take $N=F_{12}=233$ as a
representative value.  We have set $V_A=4.0$, $V_B=4.8$ and $J=1$ as
typical values corresponding to centers with similar characteristics.
The width of the instrumental resolution was $\alpha=0.15$.  The maximum
integration time and the integration time step were $50$ and $10^{-3}$,
respectively.  Smaller time steps lead essentially to the same results.

In the pure $A$ lattice the spectrum is a single Lorentzian line
centered at $E=V_A-2J=2.0$, with our choice of parameters.  When
$B$-centers are introduced according to the Fibonacci rules, a shift of
the position of the main line towards higher energies is observed.  This
shift is also observed in the case of random lattices \cite{prb1,prb2}.
The main line is now located around $E=2.30$, as shown in
Fig.~\ref{fig1}.  Besides this main absorption line, two satellite lines
can be also observed in the high-energy part of the spectrum, centered
at energies $E^{(1)}=5.92$ and $E^{(2)}=4.03$.  According to the
renormalization group techniques developed in Ref.~\cite{OPTFIBO}, none
of these satellites can be explained within the so called two-center
approach. Thus we are led to the conclusion that they are due to
long-range effects of the quasiperiodic ordering of the lattice. This
will be more evident after performing the analytical treatment.

\section{Analytical Results}

In this section we develop a theoretical approach to explain the shift
of the main line and the occurrence of well-defined satellite lines in
the high-energy region when $B$-centers are introduced quasiperiodically
in the lattice.  To this end we rewrite the system Hamiltonian ${\cal
H}={\cal H}_{\text{A}} +{\cal H}_{\text{AB}}$, where
\begin{mathletters}
\begin{eqnarray}
{\cal H}_{\text{A}} &=& V_A\sum_{n}\>a_{n}^{\dag}a_{n}
-J\sum_{n} \> (a_{n}^{\dag}a_{n+1} + a_{n+1}^{\dag}a_{n}),
\label{A1} \\
{\cal H}_{\text{AB}} &=& \sum_{n} \> (V_n-V_A) a_{n}^{\dag}a_{n}.
\label{AB1}
\end{eqnarray}
\end{mathletters}
Here ${\cal H}_{\text A}$ is the Hamiltonian corresponding to the pure
$A$ lattice.

As a first step we consider the pure $A$ lattice whose Hamiltonian is
${\cal H}_{\text{A}}$.  This Hamiltonian with periodic boundary
conditions can be exactly diagonalized, yielding the eigenvectors
\begin{mathletters}
\label{dipole}
\begin{equation}
|k\rangle=\left({1\over N}\right)^{1/2} \sum_{n}
\exp \left( 2\pi i{nk\over N} \right) a_{n}^{\dag}|\mbox{vac}\rangle,
\label{vectors}
\end{equation}
with $k=0,\ldots,N-1$. The corresponding eigenvalues read
\begin{equation}
E_{k}=V_A-2J\cos\left(2\pi{ k\over N}\right).
\label{values}
\end{equation}
\end{mathletters}
Therefore, the dipole moment matrix elements within this approximation
satisfy the relation $\langle k|{\cal D}|\mbox{vac}\rangle=\delta_{k0}$.
Consequently, one obtains that $I(E)=(1/N)\delta_{\alpha}(E-E_0)$.  This
result shows that the pure $A$ lattice presents a single Lorentzian
absorption line centered at $E_0=V_A-2J$, as we have mentioned above.

To develop an approach able to explain the results in Fibonacci systems
one must explicitly consider the effects of the topology through the
term ${\cal H}_{\text AB}$.  To carry out such an approximation we have
used the nondegenerate perturbation theory.  Let $|k^{\text
(AB)}\rangle$ the perturbed eigenvector describing the exciton state
when the influence of the term ${\cal H}_{\text AB}$ is taken into
account.  To evaluate the perturbed dipole moment matrix element, we
expand $|k^{(AB)}\rangle$ in the basis of the unperturbed eigenvectors
$|k\rangle$ given in (\ref{vectors})
\begin{mathletters}
\label{per}
\begin{equation}
|k^{\text (AB)}\rangle=|k\rangle+ \sum_{l\neq k}{\langle l|{\cal
H}_{\text AB} |k \rangle \over E_{k}-E_{l}}|l\rangle,
\label{pervector}
\end{equation}
whereas the perturbed eigenvalues are given by
\begin{equation}
E_{k}^{\text (AB)}=E_{k}+\langle k|{\cal H}_{\text AB}| k \rangle=
E_{0}^{(AB)}+4J\sin^{2}\left(\pi{k\over N}\right),
\label{pervalue}
\end{equation}
\end{mathletters}
where $E_{0}^{(AB)}=cV_B+(1-c)V_A-2J$.  Notice that ${\cal H}_{\text
AB}$ shifts all levels by the same amount.  This shift is given by
$\langle k|{\cal H}_{\text AB}| k \rangle=c(V_B-V_A)$ for all $k$.  With
the chosen parameters it amounts $0.30$, in perfect agreement with our
numerical results shown in Fig.~\ref{fig1}.

Inserting (\ref{per}) in (\ref{primero}) one finally obtains
\begin{equation}
I(E)={1\over N}\delta_{\alpha}(E-E_{0}^{\text (AB)})+
{1\over N}\sum_{k=1}^{N-1}\> |F(k)|^{2} \delta_{\alpha}(E-E_{k}^{\text
(AB)}),
\label{finalporfin}
\end{equation}
where for brevity we have defined
\begin{equation}
F(k)\equiv\frac{\langle 0|{\cal H}_{\text AB}| k \rangle}
{4J\sin^{2}(\pi k/N)}.
\label{f_def}
\end{equation}

Therefore, the spectrum consists of a main Lorentzian line centered at
$E_{0}^{(AB)}=cV_B+(1-c)V_A-2J=2.30$, in agreement with the virtual
crystal approximation, and several satellite lines at $E_{k}^{(AB)}$.
The most remarkable fact of expression (\ref{finalporfin}), however, is
that it relates the finer details of the absorption spectrum, an
experimentally measurable magnitude, with the Fourier transform of the
lattice describing the topological long-range ordering of optical
centers.  To demonstrate this point, we recall that unperturbed
eigenvectors are orthogonal.  Thus
\begin{eqnarray}
\langle 0|{\cal H}_{\text AB}| k \rangle &=& \langle 0 |\sum_{n} \>
(V_n-V_A) a_{n}^{\dag}a_{n}|k\rangle \nonumber \\
&=&{1\over N} \sum_{n} V_n \exp\left( 2\pi i {kn\over N} \right),
\ \ \ \ \ \ k\neq 0.
\label{final}
\end{eqnarray}
which is nothing but the aforementioned Fourier transform.  A
comparison of the perturbative prediction (\ref{finalporfin}) with the
numerical results is given in Fig.~\ref{fig1}, where an excellent
agreement is achieved in the weak-perturbation limit we have considered.
In particular, our analytical approach predicts the existence of two
satellite lines at $E^{(1)}=5.78$ and $E^{(2)}=4.13$

In order to get a deeper insight into the problem, we relate our results
with those reported in the various works dealing with the Fourier
transform of the Fibonacci sequence, some of them in connection with
X-ray diffraction in aperiodic semiconductor superlattices
\cite{Merlin,Godreche}.  The Fibonacci lattice is known to exhibit a
Fourier spectrum displaying well-defined Bragg peaks.  The Fourier
intensity $|\langle 0|{\cal H}_{\text AB}|k\rangle|^{2}$ consists of a
series of peaks located at values of the momentum $2\pi(k/N)$ of the
form $\tau^p$, where $p$ is an arbitrary integer.  Each peak of the
Fourier intensity leads to a large contribution to the sum
(\ref{finalporfin}) and, consequently, to a well-defined satellite line
in the absorption spectrum.

\begin{figure}

\centerline{\epsfxsize=8.0truecm \epsffile{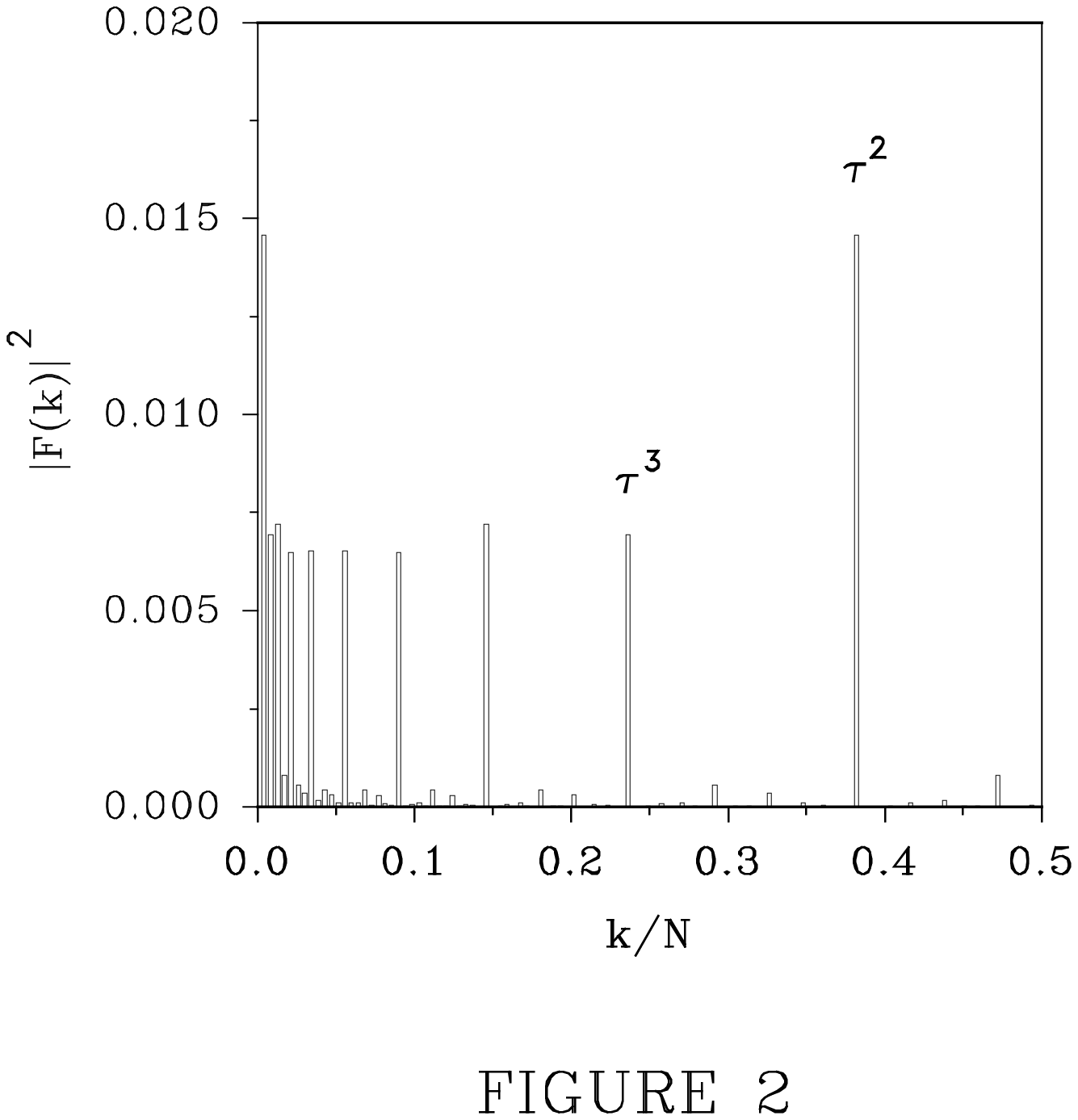}}

\caption{Plot of $|F(k)|^{2}$ as a function of the ratio $k/N$
for Fibonacci lattices, showing the ocurrence of well-defined Bragg
peaks, labelled according to successive powers of $\tau$.}

\label{fig2}

\end{figure}

Figure~\ref{fig2} shows $|F(k)|^2$ as a function of the ratio $k/N$
evaluated from expression (\ref{f_def}), the results being independent
of the system size.  A sequence of Bragg peaks, arranged according to
successive powers of $\tau$, are clearly observed in this plot.  The
stronger one, which is responsible of the highest-energy satellite line
centered at $E^{(1)}=5.78$, is located at $k_p/N \simeq 0.38\simeq
\tau^2$ ($p=2$).  Therefore the mode is located close to the top of the
excitonic band ($k/N=0.5$).  On decreasing the momentum, the next Bragg
peak is located at $k_p/N \simeq 0.23 \simeq \tau^3$ ($p=3$), leading to
an energy value $4.13$, in perfect agreement with the energy of the
second satellite peak in the high-energy region of the spectrum observed
in Fig.~\ref{fig1}.  The analytical treatment predicts the existence of
other satellite lines with smaller energy, but they are hidden by the
main absorption line.

Our results are suitable for a direct physical interpretation.  Each
satellite line of high energy is caused by the coupling of two modes,
namely the lowest-lying and $k_p$, through the topology of the
quasiperiodic lattice.  Notice that different arrangement of centers
would lead to a completely different Fourier intensity $|\langle 0|
{\cal H}_{\text AB}|k\rangle|^{2}$.  Thus, the exciton acts as a probe
of the long-range order of the quasiperiodic lattice.  This is one of
the main results of the present work since it provides us with a useful
method to be applied in experimental situations.

\section{Conclusions}

In summary, we have studied numerically the absorption spectra
corresponding to the Frenkel-exciton Hamiltonian on self-similar
quasiperiodic Fibonacci lattices.  Besides the main line, which is
shifted respect to that of the pure lattice, we found several satellite
lines in the high-energy region of the absorption spectra.  We have also
developed an analytical method which successfully explains not only the
shift of the main line and the position of the satellite lines but also
their shape.  Our analysis clearly indicates that each satellite line is
directly related to a particular Bragg peak of the Fourier transform of
the underlying lattice.  We have realized that each satellite line comes
from the coupling between the lowest-lying mode and a particular mode
lying close to the top of the excitonic band.  This relationship surely
should facilitate future experimental work on optical properties of
molecular systems with long-range order.

\acknowledgments

The author thanks E.\ Maci\'{a}, A.\ S\'{a}nchez and V.\ Malyshev for
helpful comments.  This work is supported by CICYT through project
MAT95-0325.

\end{document}